\documentclass{aa}  
\usepackage{graphicx}
\usepackage{txfonts}
\usepackage{lipsum}
\usepackage{subcaption}             
\usepackage{lscape}             
\usepackage{placeins}                                           
\usepackage{colortbl}

\def\mbh{M_{\bullet}}

\def\teff{T_{\rm eff}}
\def\rin{r_{\rm in}}
\def\rg{r_{\rm g}}

\def\sigB{\sigma_{\rm B}}

\def\re{R_{\rm E}}
\def\hp{h}
\def\aSMBBH{a_{\rm B}}

\def\msun{M_{\odot}}
\def\kms{\rm km~s^{-1}}

\def\be{\begin{equation}}
\def\ee{\end{equation}}
\def\bea{\begin{eqnarray}}
\def\eea{\end{eqnarray}}

\begin{document}

 \title{Multiwavelength periodic microlensing signatures of macrolensed supermassive binary black holes}
\titlerunning{Multiwavelength periodic microlensing signatures of  SMBBHs}
\authorrunning{Yan \& Lu}

   \author{Changshuo Yan\inst{1,2}\fnmsep\thanks{yancs@nao.cas.cn}
        \and Youjun Lu\inst{1,2}\fnmsep\thanks{luyj@nao.cas.cn}
        }

   \institute{National Astronomical Observatories, Chinese Academy of Sciences, Beijing, 100101, China\\
                 \and School of Astronomy and Space Science, University of Chinese Academy of Sciences, Beijing 100049, China\\ }

   \date{Received September 30, 20XX}

\abstract
{}
{The microlensing of lensed quasars presents a promising avenue for understanding the structure of accretion disks around supermassive binary black holes (SMBBHs). We investigated the microlensing signatures in multiband (optical, UV, and X-ray) light curves of active SMBBH systems, focusing on how these signatures depend on the mass ratio, separation, and accretion rate.} 
{We analyzed the periodic fluctuations in microlensing light curves induced by the orbital motion of SMBBHs. We examined the relation between the mass ratio and the period of variations in light curves across optical, UV, and X-ray bands.}
{We find that the periodic fluctuations in the light curves depend on the mass ratio of the black holes: for nearly equal masses, variations occur at half the orbital period, whereas for low mass ratios, the period corresponds to the orbital period influenced by the secondary mini-disk. Furthermore, all optical, UV, and X-ray light curves exhibit the same period and phase, but the amplitude of variation is greater in the UV and X-ray bands than in the optical bands. These light curves provide insights into the motion and radiation regions of the disks through wavelength-dependent periodic variations, although they yield limited constraints on the system's black hole mass or Eddington ratio, which can instead be derived from the spectral energy distribution (SED). Integrating microlensing data with SED observations is crucial for accurately constraining the parameters of SMBBH systems.}
{}

\maketitle
\nolinenumbers

\section{Introduction}
\label{sec:intro}

The $\Lambda$ cold dark matter cosmological model predicts that galaxies form hierarchically, and consequently supermassive binary black holes (SMBBHs) are expected to emerge in galactic centers following galaxy mergers, provided each galaxy harbors a central supermassive black hole \citep[SMBH; e.g.,][]{BBR80, Yu02, 2020ApJ...897...86C}. In scenarios involving gas-rich galactic mergers, close SMBBHs are likely to be encircled by a circumbinary disk with a central cavity carved out by the second SMBH. Within this cavity, each component may be surrounded by a mini-disk, fed by the circumbinary disk through gas streams, and these two mini-disks rotate around each other \citep{AL94, Ivanov99, Escala05, Hayasaki07, Hayasaki08, Dotti07, Cuadra09, DOrazio12, Farris14, Dittmann2023, Siwek23}. Such an SMBBH-triple disk system can induce various electromagnetic signatures that can be used to identify the binary nature of the system (e.g., see \citealt{Lu2025} for an overview).

The spectral energy distribution (SED) of the SMBBH-triple disk system is likely characterized by a sharp drop-off and flux deficit in the UV/optical bands, which if present would be due to its distinct disk accretion structure \citep[e.g.,][]{GM12,Sesana12,Hayasaki13,Roedig14,Yan2015,Farris15, Zheng2016,Krauth23}. While this distinct spectral feature serves as a valuable diagnostic baseline, the true SED in realistic environments is subject to substantial theoretical uncertainties. Recent 3D hydrodynamical and general relativistic magnetohydrodynamics simulations indicate that emission from shock-heated gas streams can partially fill this spectral gap, though the extent of this effect is highly variable and depends sensitively on binary parameters and gas thermodynamics \citep[e.g.,][]{Farris15, Cocchiararo24, Franchini24}. Furthermore, systems accreting at lower rates may. As these dynamic and thermodynamic complexities make it challenging to robustly extract binary parameters solely from spectral features, integrating them with independent observational probes is essential.

Microlensing amplification of lensed quasi-stellar objects (QSOs) has proven to be a powerful tool for studying the structure of accretion disks, as the Einstein radii of stars in the foreground lensing galaxies are comparable in scale to the accretion disks in the background QSOs \citep[e.g.,][]{WPS90, WMS95, Wambsganss06}. This method has been applied to measure the disk sizes for many lensed QSOs through long-term monitoring of their variations in the UV/optical and X-ray bands, providing crucial insights into the disk structure and underlying physics \citep[e.g.,][]{Morgan12, Chartas2017, Guerras2017, Cornachione2020}. \citet{Yan2014} have demonstrated that the microlensing can also resolve the unique disk structure of lensed SMBBH-triple disk systems via periodic signatures in the microlensing light curves and/or the disk size versus effective wavelength relation. \citet{Millon2022} report periodic oscillations in the $15$-year optical light curve of a lensed quasar, Q J0158-4325, which may be explained by the microlensing effect of an SMBBH system. Their results suggest that a thorough examination of the SMBBH explanation for the light curve of Q J0158-4325 is necessary, and a further detailed investigation of the microlensing effects of the SMBBH-triple disk system in multiple bands is of great importance. Such an investigation could provide a cross-check for the SMBBH interpretation.

In this study we investigated the optical, UV, and X-ray band microlensing light curves of active SMBBH systems with varying mass ratios and accretion rates. The paper is organized as follows. In Sect.~\ref{sec:model} we briefly describe the simple model for the active SMBBH-triple disk system used to calculate the surface brightness distribution of the system in the different bands and the SED. In Sect.~\ref{sec:results} we present the main results for the UV/optical and X-ray band microlensing light curves and SEDs for SMBBH-triple disk systems with different physical parameters. The effects of the streams that connect the outer circumbinary disk with the inner two mini-disks are discussed in Sect.~\ref{sec:dis}. Our main conclusions are summarized in Sect.~\ref{sec:con}.

\section{Models}
\label{sec:model}

We introduce a simple SMBBH-triple disk model in this section. We first describe the geometry and disk structure of the system, then the method for calculating SED and the light curves of the system in the UV/optical and X-ray bands due to microlensing effects. 

\subsection{Geometry and disk structure}
\label{sec:disk}

Suppose an SMBBH system with the primary and secondary component masses of $M_{\bullet,1}$ and $M_{\bullet,2}$, respectively, and a semimajor axis of $\aSMBBH$. We assume that a massive disk initially rotates around the primary SMBH. The secondary SMBH gradually migrates into the disk and carves out a gap or even a cavity within a radius $\approx2\aSMBBH$ from the center of mass. Outside this gap or cavity, a circumbinary disk persists, while each SMBH possesses its own accretion disk within the gap or cavity, fed by the circumbinary disk via gas streams \citep{Farris14, DOrazio12, Gold2019, Paschalidis2021, Dittmann2023, Siwek23}.

The outer boundary of the accretion disk around each SMBH can be constrained by the Roche lobe radius \citep{Eggleton},
\begin{equation}
R_{\rm RL}(x)=\frac{0.49 x^{2/3} \aSMBBH}{0.6x^{2/3}+\ln(1+x^{1/3})},
\label{eq:roche}
\end{equation}
where $x$ is the mass ratio ($q$ or $1/q$). We adopted the Roche lobe radius ($R_{\rm RL}$) as an upper limit for the mini-disk outer boundaries. In reality, tidal forces \citep{Paczynski77, Roedig14} and disk-binary interactions \citep{Farris14} typically truncate these disks at slightly smaller radii. While $R_{\rm RL}$ provides a practical analytical baseline, adopting a smaller, more realistic truncation radius would yield a more compact emission region, thereby further enhancing the amplitude of the microlensing-induced periodic peaks. For simplicity, we assumed that the SMBBH systems are on circular orbits, and the SMBBH and its associated triple accretion disks are coplanar.

We adopted the standard relativistic thin disk model (\citealt{NT73}; see also \citealt{SS73} for a nonrelativistic one) to describe the accretion flow around each SMBH as well as the circumbinary disk. According to this model, the disk emission can be approximated as multicolor black body radiation, with an effective temperature at any given radius ($r$) as
\bea
\teff(r) & = & \left[\frac{3G\mbh \dot{M}_{\rm acc}}{8\pi\sigB r^3}\left(1-\sqrt{\frac{\rin}{r}}\right)\right]^{1/4} \nonumber \\
& \simeq & 2\times 10^5~{\rm K} \left(\frac{0.1}{\epsilon} \right)^{1/4} \left(\frac{f_{\rm E}}{0.3}\right)^{1/4} \left(\frac{10^8\msun}{M_{\bullet}}\right)^{1/2} \nonumber \\
& & \times \left(\frac{\rin}{r}\right)^{3/4} \left(1-\sqrt{\frac{\rin}{r}}\right)^{1/4}.
\label{eq:temp}
\eea
Here $G$ is the gravitational constant, $\sigB$ is the Stefan-Boltzmann constant, $\mbh$ and $\dot{M}_{\rm acc}$ are the SMBH mass and the accretion rate, $\rin$ represent the inner edge of the disk, $\epsilon$ is the radiative efficiency, and $f_{\rm E}$ is the Eddington ratio. Observations suggest that $\epsilon \approx 0.1$ \citep[e.g.,][]{YT02, YL08, Shankar13}, corresponding to an SMBH spin of $\approx 0.67$, and then the inner disk radius $\rin = r_{\rm ISCO} \approx 3.5\rg$, where $\rg = GM_{\bullet}/c^2$ and $r_{\rm ISCO}$ are the gravitational radius and the radius of the innermost stable circular orbit of the SMBH.

The monochromatic specific intensity at wavelength $\lambda$ and radius $r$ is given by
\begin{equation}
B_{\lambda}(r)=\frac{2 hc^2/\lambda^5}{\exp\left[\frac{\hp c}{\lambda k_{\rm B}\teff(r)}\right]-1},
\end{equation}
where $\hp$ is the Planck constant, $c$ is the speed of light, and $k_{\rm B}$ is the Boltzmann constant. 
We defined the surface brightness for wavelength ($\lambda$) as
\begin{equation}
I_{\lambda}(r)=\int_{\lambda-\Delta\lambda}^{\lambda+\Delta\lambda} B_{\lambda'}(r)d\lambda'.
\end{equation}

\noindent where we take $\Delta\lambda=0.03\lambda$. In a more realistic situations, the range of specific observation bands and the influence of the transmission curve will be considered here.

For an SMBBH-triple disk system, the temperature profiles of the disks are assumed to follow that given by the standard thin disk model (Eq.~\ref{eq:temp}). Each disk is truncated at its respective inner and outer boundaries. The primary and secondary SMBH disks have outer radii $r_{\rm out,1}=R_{\rm RL}(1/q)$ and $r_{\rm out,2}=R_{\rm RL}(q)$, respectively, and inner radii at $r_{\rm in,1}=3.5GM_{\bullet,1}/c^2$ and $r_{\rm in,2}=3.5GM_{\bullet,2}/c^2$. The circumbinary disk is truncated at an inner radius $r_{\rm in}= 2\aSMBBH$ and an outer radius $r_{\rm out}=1000\rin$, beyond which the cooler disk emission contributes negligibly to our optical and X-ray bands. The gas streams connecting the outer circumbinary disk with the inner two mini-disks are ignored if not otherwise stated.

The Eddington ratios for disk accretion onto the two SMBHs are $f_{\rm E,1}$, $f_{\rm E,2}$, and for the circumbinary disk, $f_{\rm E,c}=f_{\rm E,1}/(1+q)+qf_{\rm E,2}/(1+q)$, which are derived by assuming continuous accretion $\dot{M}=\dot{M}_{\bullet,1}+\dot{M}_{\bullet,2}$, note that  $\dot{M}_{\bullet,j}=f_{{\rm E},j} \dot{M}_{{\rm Edd},j} =f_{{\rm E},j} \frac{4\pi G m_{\rm H}M_{\bullet,j}}{c\sigma_T\epsilon}$, with $m_{\rm H}$, and $\sigma_{\rm T}$ representing the mass of a hydrogen atom and the cross section of Thompson scattering.

By utilizing the simple single or triple-disk model outlined above, we can derive the UV/optical band brightness distribution for an active SMBBH-triple disk system. For comparison, we also considered the brightness distribution for an active single SMBH system by using the standard thin disk model. 

For the X-ray band, we assumed that the emission is uniform and centered on the SMBH with a typical size of $20r_{\rm g,j}=20GM_{\bullet,j}/c^2$, where $j$ denotes the specific SMBH in question. This size is consistent with previous studies of the corona, which suggest that the X-ray emitting region is less than $20r_{\rm g}$ \citep[e.g.,][]{Morgan08, Dai10, MacLeod15}.

We adopted the empirical determined relation between the X-ray luminosity and UV luminosity of QSOs to estimate the X-ray luminosity by using the UV luminosities given by the accretion model. We utilized the most recent determination of $\alpha_{\rm{ox}}$ given by \cite{Gupta24}, which is luminosity-dependent, and thus we have $\log L_{\nu}(2\,{\rm keV}) = 0.85\log L_{ \nu}(2500{\rm\AA}) + 0.985$. We computed $L_{\nu}(2500{\rm\AA})$ for each mini-disk and then derived its respective X-ray luminosity. By dividing these luminosities by the respective areas, we obtained the surface brightness for both mini-disks.

\begin{table*}[ht!]
\caption {Parameters for different systems.}
\label{tab:t1}
\centering
\begin{tabular}{cccccccccccccc}
\hline\hline             
Model & $M_{\bullet}(\msun)$ & $q$ & $f_{\rm E,1}$& $f_{\rm E,2}$ & $f_{\rm E,c}$ &$r_{\rm out,1}/R_{\rm E}$ &$r_{\rm out, 2}/R_{\rm E}$&$\aSMBBH(\rg)$& $\aSMBBH/R_{\rm E}$& $P_{\rm orb}$(day)&$P_{\rm orb}(\re/v_e)$&$t_{\rm coal}$(yr) \\
\hline
S0 & $10^9$ &           &        &      & 0.1 &&& & \\
B1 & $10^9$ &     0.1 & 0.1    & 0.1   & 0.1 & 0.13 &0.044&50&0.22 &127&0.026& 231\\
B2 & $10^9$ &     0.1 & 0. 018 & 0.92  & 0.1 & 0.13&0.044&50& 0.22&127&0.026&231\\
B3 & $10^9$ &     0.5 & 0.053  & 0.19  & 0.1 & 0.096&0.071&50& 0.22&127 &0.026&231\\
B4 & $10^9$ &     1.0 & 0.1    & 0.1  & 0.1 & 0.082&0.082&50&0.22 &127 &0.026&231\\
B5 & $10^9$ &     1.0 & 0.1    & 0.1  & 0.1 &0.16 &0.16&100&0.43 &358 &0.073&1219\\
B6 & $10^9$ &     1.0 & 0.1    & 0.1  & 0.1 &0.33 &0.33&200&0.87 &1013 &0.21&19507\\
B7 & $10^8$ &     1.0 & 0.3    & 0.3  & 0.3 & 0.033&0.033&200&0.087 &101 &0.021&1951\\
B8 & $10^8$ &     1.0 & 0.3    & 0.3  & 0.3 & 0.066&0.066&400&0.17 &286 &0.058&31212\\\hline\hline
\end{tabular}
\tablefoot{Basic parameters that define the example single/triple disk systems. For the S0 system, $M_{\bullet}$ and $f_{\rm E,c}$ represent the mass of the central massive black hole and Eddington ratio of the accretion disk, respectively; for other SMBBH systems, $M_{\bullet}$, $q$, $f_{\rm E,1}$, $f_{\rm E,2}$, $f_{\rm E,c}$, $r_{\rm out,1}$, $r_{\rm out,2}$ and $a_{\rm SMBBH}$ represent the total mass, the mass ratio, the Eddington ratios of the disks associated with the primary component, secondary component, the circumbinary disk of each SMBBHs, the size of the primary disk, the secondary disk and the semimajor axis, respectively. The $P_{\rm orb}$ is the orbit period time of the system while $t_{\rm coal}=\frac{5}{256} \frac{c^5 a_{\rm SMBBH}^4}{G^3 \mbh (M_{\bullet,1}M_{\bullet,2})}$ is the coalescence time of the SMBBH due to the gravitational wave decay.}
\end{table*}

\subsection{Spectral energy distribution}

Due to the unique triple-disk structure of the active SMBBH system, its SED may differ from that of a single active SMBH system, mainly characterized by the deficiency in the optical and UV bands. The emission from the circumbinary disk and the two inner mini-disk associated with the primary and secondary SMBHs can be calculated as
\begin{equation}
F_\lambda=\frac{2 \pi \cos {\iota}}{D_L^2} \int_{r_{\rm in}}^{r_{\text {out }}} B_\lambda(r) r \mathrm{~d} r,
\end{equation}
where $D_L$ is the luminosity distance of the source, $i$ is the inclination angle of the disk to the observer's line of sight. For illustration, here we only considered the face-on case fixed $\cos \iota$ to unity. Combining the radiation from the three disk together, we obtained the SED of the SMBBH system as
\bea
\lambda L_{\lambda}&=&4\pi D_L^2 \lambda F_{\lambda}
=8\pi^2\lambda\left(\int^{r_{\rm out}}_{r_{\rm in}} B_{\lambda}(r)rdr+ \right. \nonumber\\
&&\left.\int^{r_{\rm out,1}}_{r_{\rm in,1}} B_{\lambda,1}(r)rdr
+\int^{r_{\rm out,2}}_{r_{\rm in,2}} B_{\lambda,2}(r)rdr\right).
\eea
Here we only considered the thin disk emission in the optical-UV bands. Although the existence of a corona will affect the radiation of the accretion disk, it does so only in the extreme UV band (wavelengths less than $0.1 \mu$m; \citealt{Czerny87}).

\begin{figure*}
\centering
\includegraphics[width=0.8\textwidth,angle=-90]{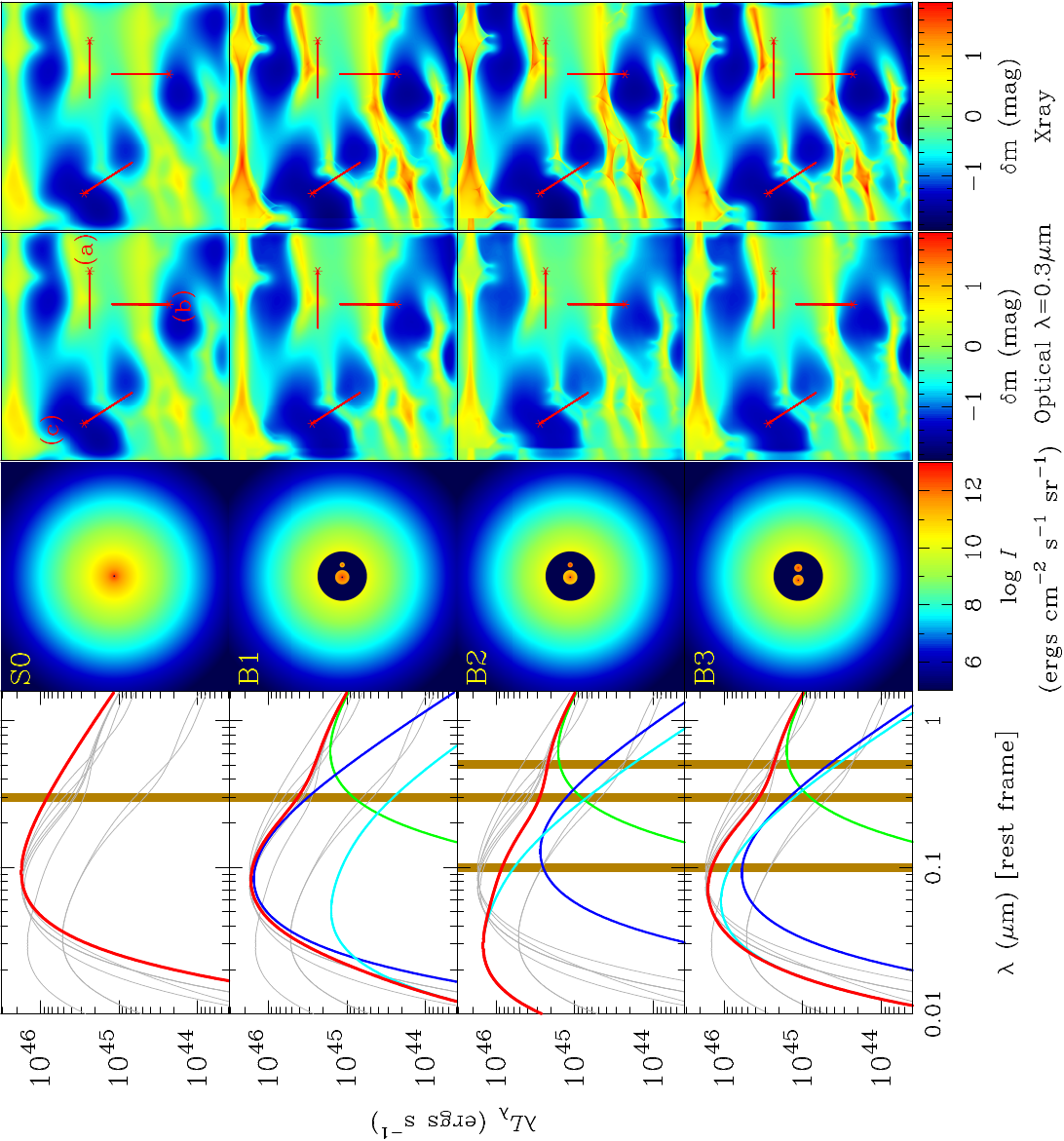}
\caption{
SEDs (first column), optical brightness distributions at a specific given time (second column), and convolved magnification maps in the optical (fourth column) and X-ray (third column) bands obtained from different models. Panels from top to bottom show the results obtained from models S0, B1, B2, and B3 (see Table~\ref{tab:t1}). In the first column, the green, blue, and cyan lines show the emissions from the circumbinary disk and the primary and secondary mini-disks, respectively, and the light gray lines show the SEDs from other systems for comparison purposes. In the second column, the optical band ($\lambda=0.3\mu$m) surface brightness distribution maps are obtained from the simple (triple) disk model described in Sect.~\ref{sec:disk} assuming a face-on view at a specific given time. The color bar at the bottom of the panels indicates the values of disk surface brightness. In the third column, the convolved magnification maps in the source planes are obtained by assuming $\lambda=0.3\mu$m-band surface brightness distributions of the sources in the corresponding second column panels. In the fourth column, the convolved magnification maps in the source planes are obtained by assuming the X-ray surface brightness is distributed homogeneously within $20R_{\rm g}$ for each mini-disk. The solid red lines depict three different relative trajectories of the system across the magnification map generated by the fixed star field, labeled as (a), (b), and (c). These relative trajectories are adopted to calculate the mock light curves shown in Fig.~\ref{fig:lightcurve}, and each asterisk represents the starting position of the SMBBH center of mass for the corresponding trajectory. Each map consists of $2048\times2048$ pixels, and the width of each map is $4$ times the Einstein radius.
}
\label{fig:f1}
\end{figure*}

\begin{figure*}
\centering
\includegraphics[width=0.6\textwidth,angle=-90]{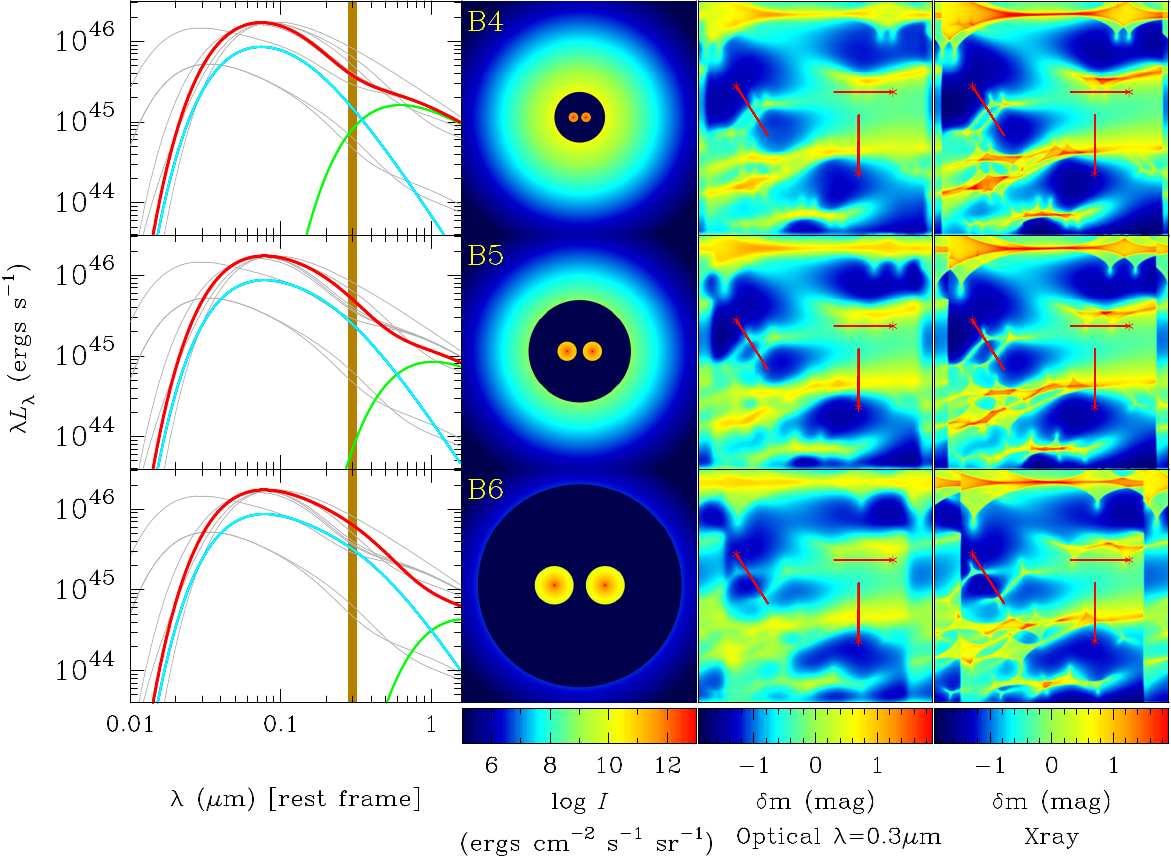}
\caption{SEDs, the optical ($\lambda=0.3\mu $m band) surface brightness map, and convolved magnification map for models B4, B5, and B6. Symbols and lines are the same as those in Fig.~\ref{fig:f1}. }
\label{fig:f2}
\end{figure*}

\begin{figure*}
\centering
\includegraphics[width=0.5\textwidth,angle=-90]{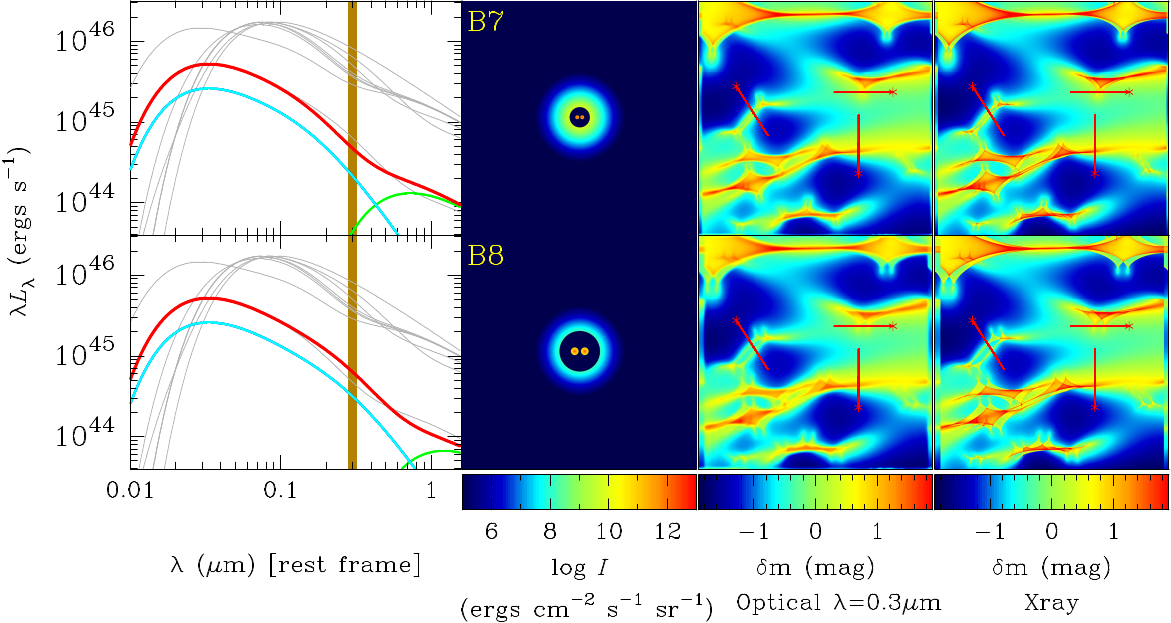}
\caption{
Same as Fig. 2 but for models B7 and B8.}
\label{fig:f3}
\end{figure*}

\subsection{Microlensing light curves in UV/optical and X-ray bands}
\label{sec:mocklc}

The magnification of a macro-image caused by microlensing is depicted through a magnification map in the source plane. Each pixel on this map represents the difference in the magnitude of the macro-image when the source is at a specific point, compared to the average macro-image magnitude. These maps are generated using the ray-shooting method, taking into account lens properties such as surface stellar mass density and shear \citep{Kochanek04, Wambsganss06, Kayser86, Paczynski86, Wambsganss90, WPS90}.

For demonstration purpose, in our simulations we used fixed mean convergence ($\kappa$) and shear ($\gamma$) values of ($0.72, 1.03$) and the stellar contribution to the convergence was set to $\kappa_*/\kappa=0.92$ (these values are the macrolensing model of image B of Q J0158-4325; see \citealt{Millon2022}). Star positions are randomly assigned, and their masses are drawn from a Salpeter initial mass function, bounded between $0.1\msun$ and $10\msun$. To generate the magnification maps, we utilized an optimized hierarchical tree code for ray-shooting \citep{Wambsganss90, WPS90}. For our specific macrolensing parameters ($\kappa_* \approx 0.662$, $\gamma=1.03$), the strong external shear highly distorts the mapping between the lens and source planes. To avoid artificial boundary effects and adequately capture the long-range deflections (diffuse flux), the code dynamically populates a much larger region than the final receiving field. Specifically, the simulation distributed $N_{\rm stars} = 16,011$ stars over a large circular region in the lens plane with a radius of $R \approx 86.5 \langle R_E \rangle$. This ensures that the deflection angles computed for the central $4 \langle R_E \rangle \times 4 \langle R_E \rangle$ magnification map are robust. The exact same random realization of the star field (identical random seed) was utilized across all models (B1-B8). Consequently, any morphological differences observed between the convolved magnification maps of different models arise strictly from the convolution process with their uniquely varying extended source geometries, not from changes in the underlying caustic network.

For the lens system, we adopted the same parameters as those for the lensing system Q J0158-4325, so the Einstein radius for a star in the source plane is $\left< \re \right> = 3.4 \times 10^{16} \left(\left< m_* \right> / 0.3 \msun \right)^{1/2}$\,cm. Quasars are extended sources at the scale of accretion disks, necessitating the convolution of original magnification maps with the surface brightness distribution maps of the sources. The SMBBHs' orbital motion introduces additional complexity. The orbital period of a SMBBH system is
\bea
P_{\rm orb}&=&2\pi \left(\frac{a^3_{\rm B}}{GM_{\bullet}}\right)^{1/2} \nonumber \\
&=&35.8\left(\frac{M_{\bullet}}{10^8\msun}\right)^{-1/2} \left(\frac{a_{\rm B}}{100 \rg}\right)^{3/2} \text{days}.
\eea
The microlensing variation timescale is determined by the source size and the effective relative transverse velocity $v_{\rm e}$ between the source and the lens. We set $v_{\rm e} = 800 \kms$, a typical source-plane effective velocity that accounts for the velocity dispersion of stars in the lensing galaxy, alongside the relative proper motions of the observer, lens, and source \citep[e.g.,][]{MK11}. We monitored the light curves over a period of $\sim \left<\re \right> / v_{\rm e} \simeq 13.5$ years. Crucially, this crossing time is measured in the observer frame, meaning it must be compared against the observed orbital period $P_{\rm obs} = P_{\rm orb}(1+z)$. For the subparsec systems modeled here, the observed orbital periods are on the order of months to a few years, ensuring that $P_{\rm obs} \ll \left<\re \right> / v_{\rm e}$. In this regime, the SMBBH system completes multiple orbits during a single caustic crossing event, allowing the orbital motion to significantly imprint periodic fluctuations onto the microlensing light curve; otherwise, there would no microlensing-induced periodicity.

The SMBBH system's surface brightness distribution evolves with time, particularly within the gap/cavity due to the rotation of the SMBBH and its disks around the center of mass. At each monitoring epoch, we convolved the current surface brightness map with the original magnification map to obtain the light curve. And then we defined a trajectory, and the ``measured" magnitude at each epoch $i$ is given by
\begin{equation}
m_i=-2.5\log \mu_{{\rm micro},i} -2.5\log |\mu_{{\rm macro},i}| + m_{\rm int},
\end{equation}
where $\mu_{{\rm micro},i}$ and $\mu_{{\rm macro},i}$ are the microlensing and macrolensing amplification factors at epoch $i$, respectively, and $m_{\rm int}$ is the intrinsic magnitude at epoch $i$. 
For lensing systems with multiple images, after considering the time delay effect and the magnification effect, one can obtain a light curve that contains only the microlensing-induced variations by subtracting the intrinsic variations of the source encoded in different images. In this study we were only concerned about the effects due to microlensing and thus the deviations of the measured magnitude ($m_i$) at each epoch ($i$) from the mean magnitude ($\langle m_i\rangle$):
\begin{equation}
\delta m_i=m_i-\left\langle m_i\right\rangle= - 2.5\log \left(\mu_{{\rm micro},i}\right).
\label{eq:deltam}
\end{equation}

\begin{figure*}
\centering
\includegraphics[width=0.6\textwidth,angle=-90]{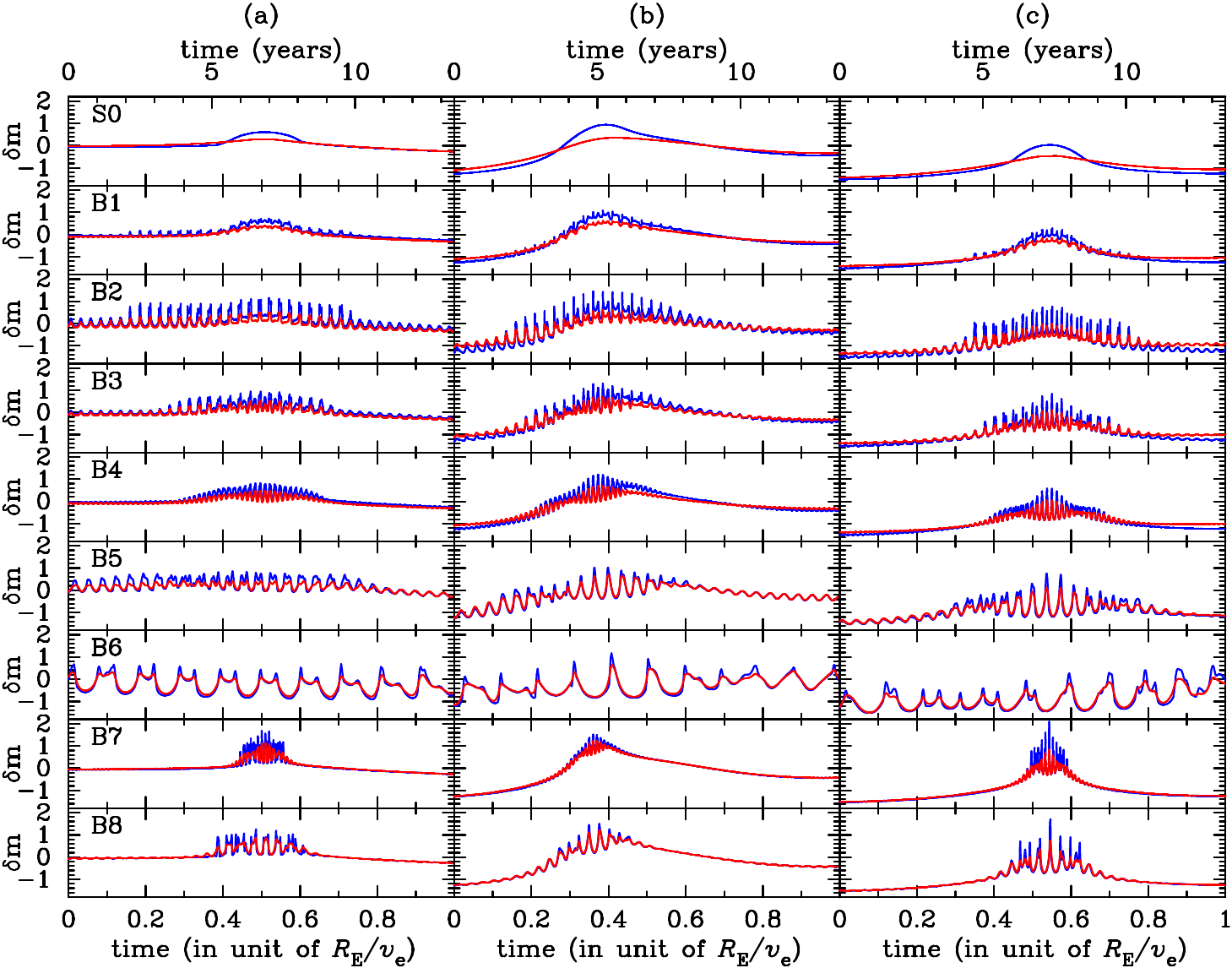}
\caption{
Mock light curves at the optical ($\lambda=0.3\mu$m) and X-ray bands obtained by adopting trajectory a (left column), trajectory b (middle column), or trajectory c (right column) for different systems. Panels from top to bottom show the results for the image B resulting from the systems S0, B1, B2, B3, B4, B5, B6, B7, and B8,. The red and blue lines represent the mock light curves of the optical and X-ray band, respectively. The gray and white stripes indicate the orbit period of each SMBBH system. The bottom $x$-axis shows the time in units of the crossing timescale $\left<\re \right> / v_{\rm e}$, while the top $x$-axis displays the corresponding physical time in years. Here, $\left<\re \right> = 3.4 \times 10^{16} \left(\left< m_* \right> / 0.3 \msun \right)^{1/2}$\,cm is the average Einstein radius of the stars in the source plane, and $v_{\rm e} = 800 \kms$ is the effective relative transverse velocity of the system. These parameters yield a characteristic crossing timescale of $\left<\re \right> / v_{\rm e} \simeq 13.5$\,years.
}
\label{fig:lightcurve}
\end{figure*}

\begin{figure*}
\centering
\includegraphics[width=0.6\textwidth,angle=-90]{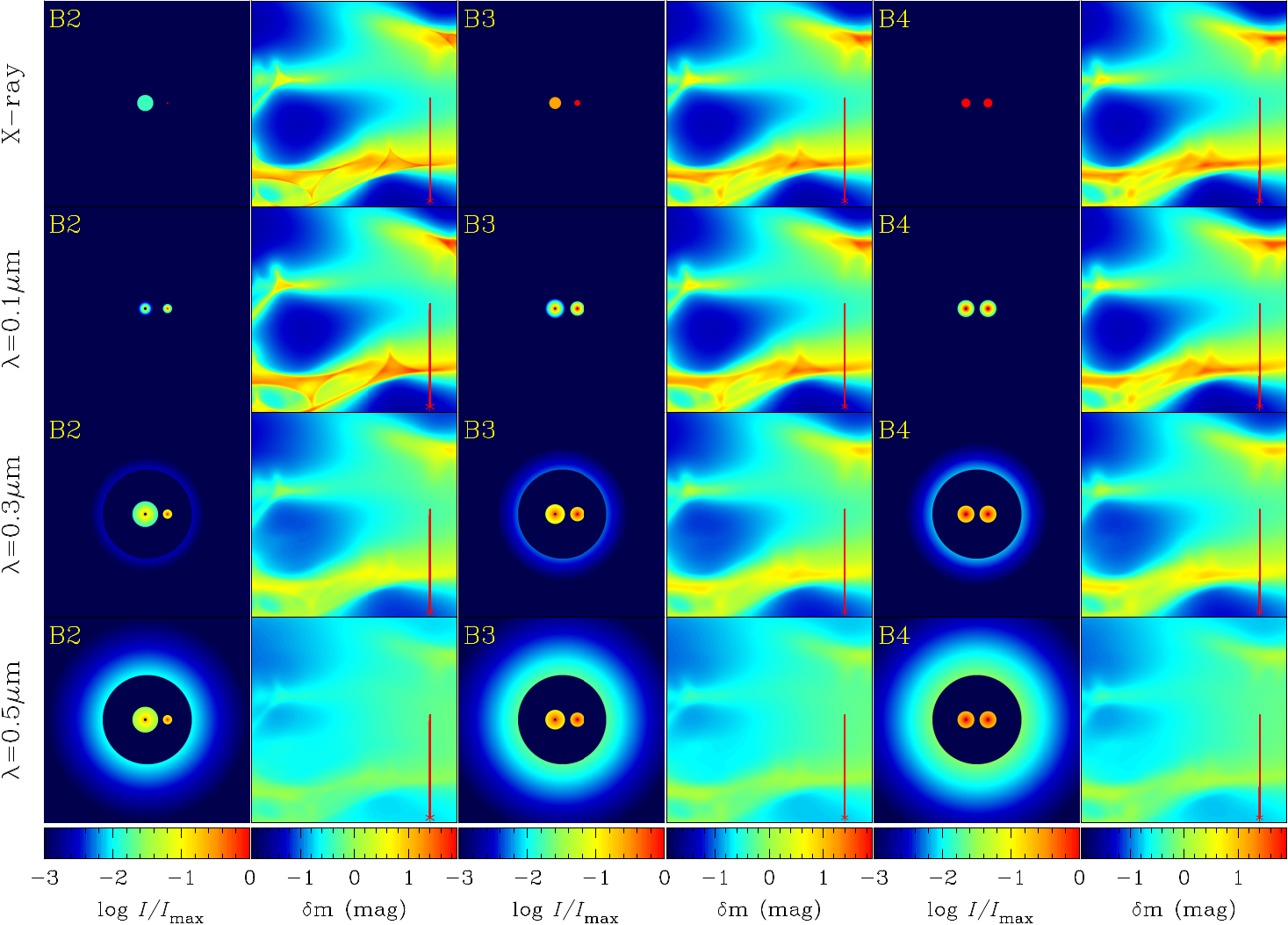}
\caption{
Relative surface brightness map and convolved magnification map for models B2, B3, and B4 in different bands. Symbols and lines are the same as those in Fig.~\ref{fig:f1}. Only the central 1024 ×1024 pixels are shown, and the width of each map is 2 times the Einstein radius.
}
\label{fig:f1b}
\end{figure*}

\section{Results}
\label{sec:results}

We employed nine sets of models, each with different parameters, to investigate the observational characteristics of microlensing light curves and SEDs of SMBBH systems. Table~\ref{tab:t1} outlines the parameter configurations for each model. Model S0 represents a single SMBH accretion system, while models B4-B8 correspond to equal mass SMBBH-triple disk system. Model B3 has a mass ratio of $0.5$, and models B1 and B2 feature SMBBHs with a mass ratio of $0.1$. We assumed continuous accretion in these models, meaning the accretion rate of the circumbinary disk equals the sum of the accretion rates of secondary and primary SMBBH components. Based on simulation results from \citet{Duffell2020}, we set $\dot{M}_{\bullet,1}=\dot{M}_{\bullet,2}(0.1+0.9q)$ for models B2-B8. For comparison, in model B1, we set $\dot{M}_{\bullet,2}=q\dot{M}_{\bullet,1}$, maintaining the  same Eddington ratio for the disks around the primary and secondary components. Models B4, B5, and B6 investigate different semimajor axes, while models B7 and B8 explore variations in total SMBH mass.

Figure~\ref{fig:f1} 
illustrates the SEDs, surface brightness distributions, and convolved magnification maps for models S0, B1, B2, and B3, respectively. In model B1 
the secondary disk (hereafter disk2) is less luminous than the primary disk (hereafter disk1), and the disk1 is located closer to the center of mass than disk2. Consequently, the convolved magnification map for B1 closely resembles that of S0 due to the minimal impact of the secondary disk. However, the presence of a gap in the SMBBH system B1's disk results in an SED significantly differs from S0.

Figure~\ref{fig:f2} presents the SEDs, surface brightness distributions, and convolved magnification maps for models B4, B5, and B6, while Fig.~\ref{fig:f3} does the same for models B7 and B8. Differences among models B4, B5, and B6 arise from varying SMBBH separations, affecting the positions of gaps in their SEDs. The convolved magnification maps reveal two sets of caustics whose separation corresponds to the distance between the disks. Larger separations yield larger Roche lobes, resulting in increased disk sizes and more diffuse convolved magnification maps in the optical band. For models B7 and B8, the SMBH masses are set to $10^8\mbh$, leading to significantly lower luminosity and smaller disk sizes compared to models B1 through B6.

Figure~\ref{fig:lightcurve} shows the simulated microlensing light curves for the trajectories (a), (b), and (c) as illustrated in the S0 magnification map in Fig.~\ref{fig:f1}. The light curves are shown in both the optical ($\lambda \approx 0.3\mu$m) and X-ray bands, with gray stripes indicating the orbital period of each SMBBH system. Each light curve exhibits an overall variation trend (a main caustic crossing peak) overlaid by periodic fluctuations. The main peak in each light curve corresponds to caustic crossing events, with the amplitude of these peaks being sensitive to the ratio of the caustic size to the size of the emitting region (including the two mini-disks and the circumbinary disk). A larger ratio results in higher peak amplitudes, explaining why X-ray band light curves show more pronounced microlensing variations compared to the optical band, as the X-ray emission region is more compact.

Compared to single disk system light curves, those from SMBBH-triple disk systems exhibit more intricate structures, notably periodic variations induced by the SMBBH's rotation. During a caustic crossing event, one disk (disk1 or disk2) initially encounters the caustic and subsequently moves away, followed by the other disk crossing the caustic approximately half an SMBBH period later. The first disk then returns to cross the caustic again after about one SMBBH period. This periodic caustic crossing continues until the caustic traverses the entire gap region, resulting in multiple peaks in the light curves (we mark the light curve peak associated with disk1(disk2) as peak1(peak2)), especially when emissions from the two disks are comparable, as seen in models B4-B8. The period of light variation is half of the $P_{\rm orb}$. 

When the source enters and exits the areas bordered by caustic lines or undergoes high magnification events, the light curve will exhibit characteristic double-horned structures \citep{Vernardos24}, which are evident in model B6's light curve (a). However, when the trajectory is away from the caustic, these small double-horned structures disappear, as seen in model B5's light curve (a) when time is greater than $0.8$.

If the mass ratio of the two SMBBH components ($q$) is small, the primary disk remains close to the SMBBH system's mass center. In such cases, the second disk crosses the caustic roughly one SMBBH period after a previous crossing, resulting in a variation period equivalent to $P_{\rm orb}$ (see cases B1 and B2). If the emission from the primary disk is negligible (case B2), the secondary disk dominates the light changes. Conversely, when the primary disk's emission is dominant, the light curves of case B1 and S0 are quite similar, with disk2's emission contributing to a small periodic change in the light curve. With a moderate mass ratio (e.g., $q = 0.5$), as seen in model B3, a prominent peak in the light curve corresponds to disk2, followed by a smaller peak from disk1. When the SMBBH is positioned away from the caustic, resulting in minimal magnification, the lower peak becomes less pronounced, leading the light curve to exhibit a single peak, similar to low mass ratio cases (see Fig.~\ref{fig:lightcurve} B3 when time is greater than 0.5 or smaller than 0.2).

\begin{figure*}
\centering
\includegraphics[width=0.8\textwidth,angle=0]{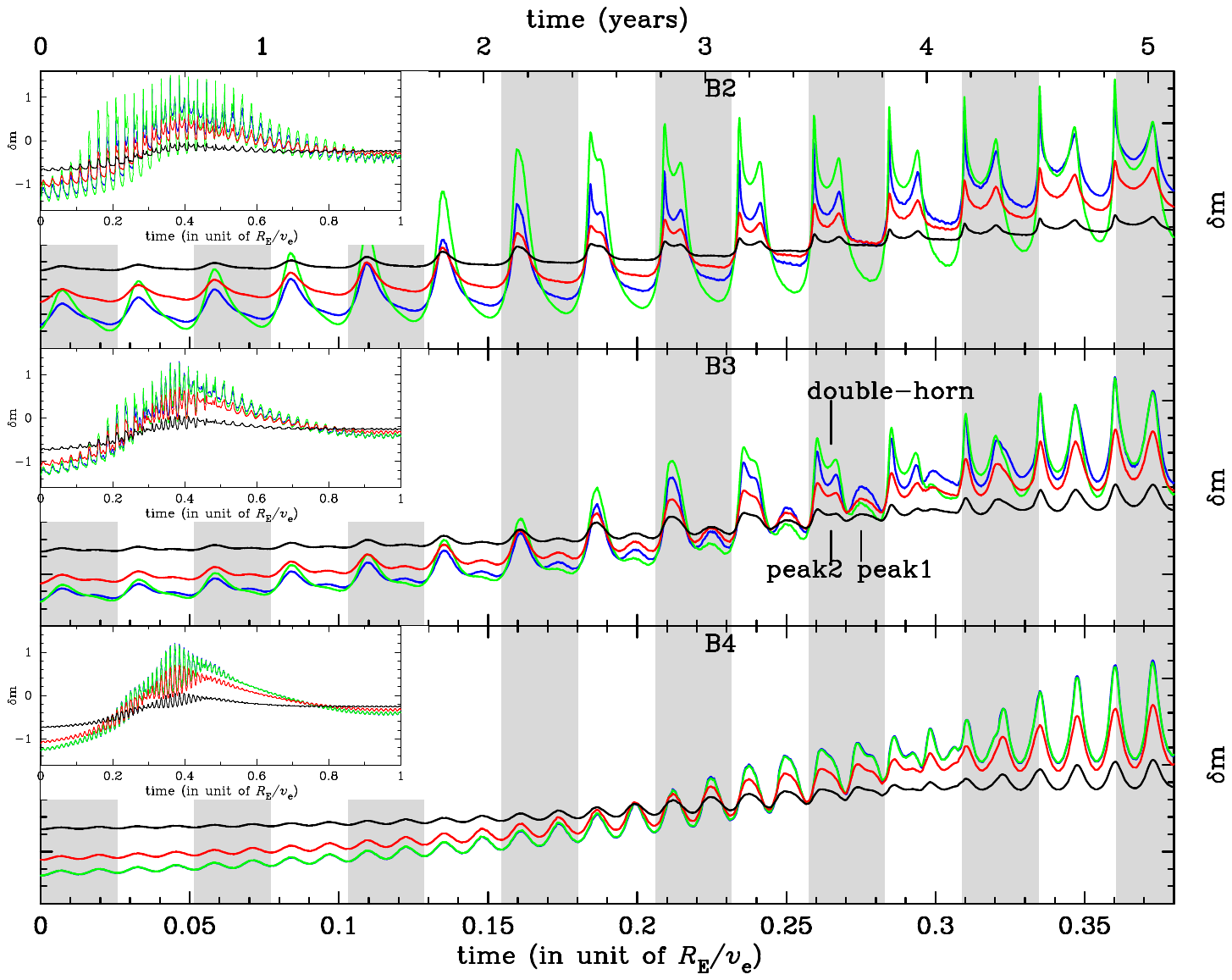}
\caption{
Mock light curves obtained by adopting the trajectory (trajectory b) in Fig.~\ref{fig:f1b} for the models B2, B3, and B4, respectively. The blue, green, red, and black lines in each panel show the mock light curves of the X-ray band and three different bands ($0.1\mu$m, $0.3\mu$m, and $0.5\mu$m), respectively. The gray stripes indicate the orbit period of each SMBBH system.The small panel in the upper-left corner shows the entire light curve, while the main panel displays a portion of it.
}
\label{fig:lightcurve2}
\end{figure*}

For models B2, B3, and B4, we explored the effects of different observation bands, such as X-ray, $0.1\mu$m, $0.3\mu$m, and $0.5\mu$m, as indicated by the yellow thick vertical lines in Fig.~\ref{fig:f1}'s left panel. 
Table~\ref{tab:t2} presents the truncated mini-disk radius and the half-light radius of these systems, the latter being the radius within which half of the total emitted light is contained. This can be determined by integrating the specific intensity at any given wavelength, $\lambda$, across the disk radii. Figure~\ref{fig:f1b} displays the corresponding surface brightness distributions and convolved magnification maps in these bands. Figure~\ref{fig:lightcurve2} presents mock light curves using the trajectory in Fig.~\ref{fig:f1b} at four different bands for models B2, B3, and B4. 
 
As illustrated in the small panels in the upper left corner of Fig.~\ref{fig:lightcurve2}, the duration of caustic crossing events—associated with the main peak—increases with longer wavelengths and magnification decreases with increasing wavelength due to the expanding size of the emission region. The X-ray and UV ($\lambda=0.1\mu$m) light curves show similar peak widths, as indicated by the blue and green lines. This similarity arises because we assumed X-ray radiation is primarily concentrated within a range of $20 r_g$ at the center of the mini-disks. As shown in Fig.~\ref{fig:f1b}, the UV radiation at a wavelength of $\lambda=0.1\mu$m also mainly originates from these two mini-disks. Furthermore, from the 5th and 6th lines of Table 2, we can see that the size of the total emission region for X-ray and UV radiation, which includes the two mini-disks, is similar. However, the emission at 0.3 $\mu$m and 0.5 $\mu$m originates partly from the circumbinary disk. Coupled with the standard accretion disk model's prediction that longer wavelengths come from larger radii, this results in a more extensive overall emission region.
 
We now examine the fine structure in the light curves. The observed small periodic peaks, which corresponding with the mini-disk crossing the caustic, are closely related to the size of the mini-disk emitting region. As shown in Fig.~\ref{fig:lightcurve2} for wavelengths of $\lambda = 0.1$, 0.3, and 0.5 $\mu$m, the magnification of these periodic peaks increases as the size of the emission region decreases. The situation is more complicated for X-rays because we adopted the empirical determined relation between the X-ray luminosity and UV luminosity of QSOs to arrange the X-ray luminosity into two mini-disks; therefore, the  Eddington ratio of the primary mini-disk  is lower, but its X-ray luminosity is relatively higher. The luminosity ratio of the secondary mini-disk to the primary mini-disk is greater in the UV band than in the X-ray bands. Consequently, the amplitude of peak1 in the X-ray light curve is larger than that in the UV, while the amplitude of peak2 in the X-ray light curve is smaller than in the UV. In summary, both the distribution of emissions within the two mini disks and the size of their emission regions can influence the magnification of periodic peaks.

\begin{table*}[ht!]
\caption {Parameters for B2, B3, and B4 in different bands.}
\label{tab:t2} 
\centering
\begin{tabular}{ccccc|cccc|cccc}
\hline\hline 
Model &&&B2&&&&B3&&&&B4&  \\
\hline
band&X-ray&0.1$\mu$m&0.3$\mu$m&0.5$\mu$m&X-ray&0.1$\mu$m&0.3$\mu$m&0.5$\mu$m&X-ray&0.1$\mu$m&0.3$\mu$m&0.5$\mu$m\\
\hline
$r_{\rm out,1}/R_{\rm E}$ &0.079&0.13&0.13&0.13&0.058&0.096&0.096&0.096&0.043&0.082&0.082&0.082\\
$r_{1/2,1}/R_{\rm E}$ &0.056&0.034&0.061&0.069&0.041&0.031&0.050&0.055&0.030&0.028&0.044&0.047\\
$r_{\rm out, 2}/R_{\rm E}$&0.008&0.044&0.044&0.044&0.029&0.071&0.071&0.071&0.043&0.082&0.082&0.082\\
$r_{1/2,2}/R_{\rm E}$ &0.006&0.014&0.023&0.024&0.023&0.023&0.037&0.039&0.030&0.028&0.044&0.047\\\hline
$\frac{r^2_{1/2,1}+r^2_{1/2,2}}{R^2_{\rm E}}$&0.0032&0.0014&0.0042&0.0053&0.0022&0.0015&0.0039&0.0045&0.0018&0.0016&0.0039&0.0044 \\
$\frac{r^2_{\rm out,1}+r^2_{\rm out,2}}{R^2_{\rm E}}$&0.0063&0.019&0.019&0.019&0.0042&0.014&0.014&0.014&0.0037&0.013&0.013&0.013\\
\hline
$L_2/L_1$&0.679&2.372&0.544&0.402&0.926&1.439&0.869&0.789&1.0&1.0&1.0&1.0\\
\hline
\end{tabular}
\tablefoot{
The mini-disk size of models B2, B3, and B4 in different bands. $r_{\rm out,1}$ and $r_{\rm out,2}$ are the mini-disk size of primary and secondary disk, respectively. 
$r_{1/2}$ is the half light radius of the truncated disk, within which half of the light is contained, and it can be estimated by integrating the specific intensity at $\lambda$ over the disk radius from $r_{{\rm in},i}$ to $r_{{\rm out},i}$, with the subscript $i$ represent the primary or secondary mini-disk. The ratio $L_2/L_1$, represents the luminosity of the secondary mini-disk relative to the primary mini-disk in the corresponding band.  
}
\end{table*}

In model B3, with a mass ratio of $q = 0.5$, the primary and secondary disks are comparable in size, allowing for the clear identification of periodic peaks associated with each disk, labeled as peak1 and peak2 in Fig.~\ref{fig:lightcurve2}. In contrast, model B2 ($q=0.1$) exhibits a single periodic peak associated with disk2 (peak2), while its primary peak(peak1) remains undetectable. In model B4 ($q=1.0$), peak1 and peak2 are identical as emissions from the two mini-disks are the same.
When the trajectory approaches the caustic, magnification increases, the light curve first revealing a double-horned structure within peak2, then right horn of the peak2 merge with the left horn of the peak1. The double-horned profile and small peaks related with mini-disks mix together.
In summary, the light curves of binary black holes display intricate and diverse characteristics, reflecting their orbital motion, the size of the black holes, and the distribution of the emitting regions.

It is important to note that our analysis assumes the accretion disk radiation follows the standard disk model. However, when the Eddington accretion rate of the primary black hole is relatively low, significantly below 0.1, the accretion flow is likely to transition to a radiatively inefficient state \citep{Yuan2014}, corresponding to a mixed accretion mode where the primary mini-disk becomes radiatively inefficient while the secondary maintains a high accretion rate \citep{Tiede2025}. In this scenario, the disk may become an advection-dominated accretion flow, resulting in significantly reduced radiation emission. Consequently, the luminosity ratio between the secondary and primary mini-disks increases, potentially rendering peak1 undetectable but peak2 remains.

Furthermore, it is important to clarify that our theoretical framework inherently accommodates the intrinsic variability of the accreting binary, as represented by the term $m_{\rm int}$ in Eq. (8). In reality, quasars exhibit intrinsic stochastic variability, typically modeled as a damped random walk, and active SMBBH systems typically exhibit intrinsic periodicities driven by accretion rate modulations, relativistic Doppler boosting, and self-lensing \citep[][see also \citealt{Lu2025}]{DOrazio2023}. To extract the pure microlensing signatures predicted in this paper, it is not necessary to adopt specific models for these complex intrinsic fluctuations for the following reason. Any intrinsic quasar variability (either stochastic or periodic variation) is imprinted identically on all macro-images, separated only by the system's macroscopic time delays. Provided these time delays can be well constrained, subtracting the time-shifted light curve of image A from image B cancels out the $m_{\rm int}$ component. Through this subtraction, the intrinsic modulations can be effectively completely removed, leaving only the uncorrelated microlensing flares and revealing the underlying caustics.

The SED of a SMBBH-triple disk system is determined by the total black hole mass, mass ratio, Eddington ratios, sizes and the location of the three disks. The left panels of Figs.~\ref{fig:f1}, \ref{fig:f2}, and \ref{fig:f3} show the SEDs resulting from the nine models with different parameters as listed in Table~\ref{tab:t1}. The total mass and Eddington ratios determine the emission from circumbinary disk and the total accretion rate of the two mini-disk if we assume the continuous accretion. The separation determines the wavelength of the deficit structure, which can be observed when comparing the SEDs of B4, B5, and B6 (see the left panels of Fig.~\ref{fig:f2}). As shown in Fig.~{\ref{fig:f1}, SEDs resulting from models B1 and B3 are similar, especially in the optical bands, but their microlensing light curves differ significantly. SMBBH models B2, B3, and B5 can produce similar periodic variations in microlensing light curves, yet the SEDs of these three models are quite different. Macrolensing only affects the overall luminosity, while microlensing can alter the shape of the SED due to different amplification rates for radiation at various wavelengths. In this analysis, we ignored gravitational lensing effects on the SED and focused solely on the SED when the source is far from the caustic or without microlensing. 
Additionally, extreme Eddington ratios can invalidate the standard thin-disk assumption, introducing greater uncertainties into the SED. For instance, super-Eddington accretion flows are better described by the slim disk approximation \citep{2006ApJ...648..523W}, whereas very low accretion rates may induce mixed accretion modes that radically alter the broadband spectrum \citep{Tiede2025}. Although the characteristic UV/optical deficit generally persists for the highly accreting mini-disks modeled here, such underlying spectral degeneracies strongly motivate the use of microlensing as an independent constraint.

In summary, the complex microlensing light curves of lensed SMBBH quasars, as seen in Figs.~\ref{fig:lightcurve} and \ref{fig:lightcurve2}, may offer valuable insights into the nature of SMBBH systems in these quasars, if present. For SMBBH systems with pronounced rotational effects, even a single-color microlensing light curve can strongly indicate the presence of SMBBHs. However, to determine system parameters, one may need to combine microlensing light curves with SED observations.

\begin{figure}
\centering
\includegraphics[width=0.58\textwidth,angle=-90]{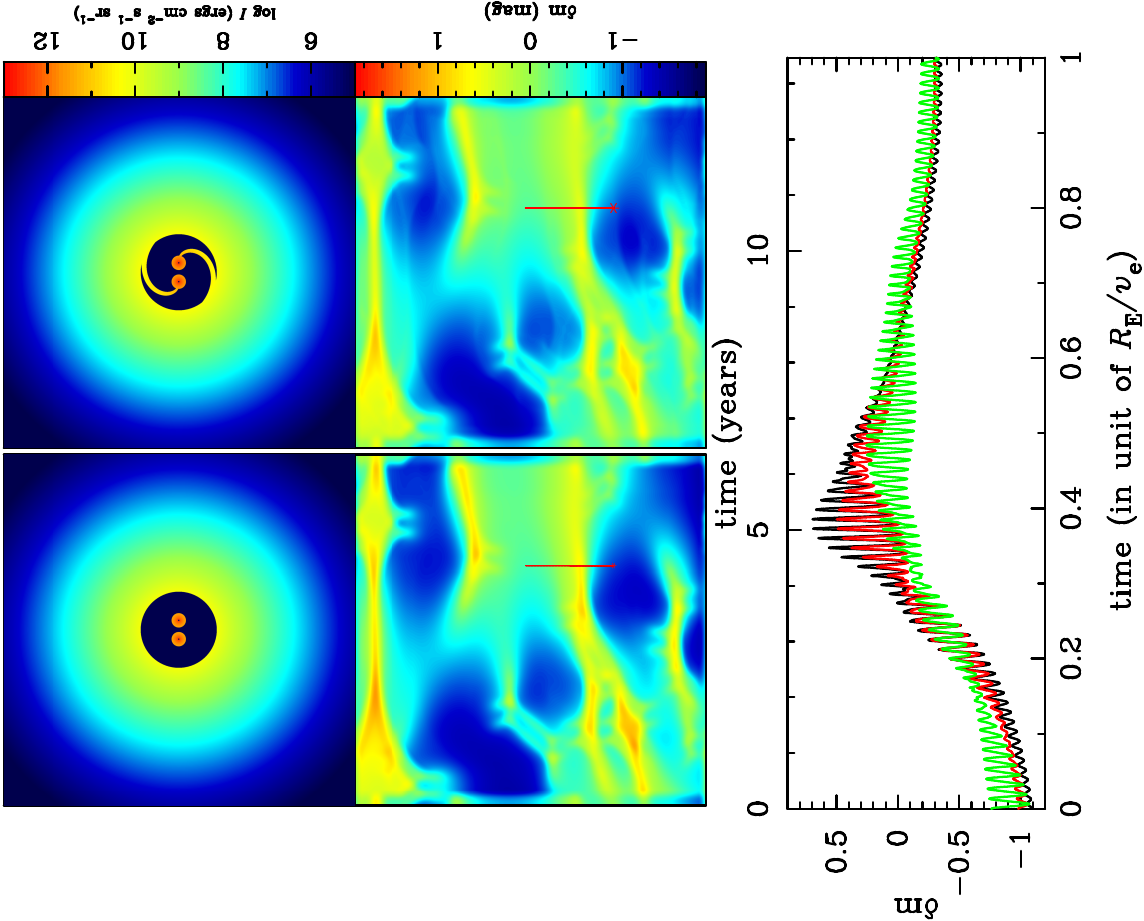}
\caption{
Top: Optical ($\lambda=0.3\mu$m band) surface brightness map and convolved magnification map for model B4 without and with streams.
Bottom: Light curves of trajectory (b) without streams (black line), light curves of the system with the optical luminosity ($\lambda=0.3\mu$m band) $L_{\rm stream}\approx 0.2 L_{\rm disks}$ (red line), and the light curves of system with $L_{\rm stream}\approx L_{\rm disks}$ (green line).
}
\label{fig:compare}
\end{figure}

\section{Impact of streams}
\label{sec:dis}

Numerical simulations of SMBBHs embedded in accretion disks have shown that there are gas streams connecting the outer circumbinary disks with the inner mini-disk around each SMBH component  \citep[e.g.,][]{AL96, DOrazio12, Siwek23}. The streams also radiate and thus may influence the microlensing light curves. Some studies suggest that the radiation from streams is an order of magnitude lower than that from disks \citep[e.g.,][]{Franchini24}, while other studies indicate that the radiation levels are comparable to those from disks \citep[e.g.,][]{Cocchiararo24}. To examine the impact of the streams on the light curves, we adopted a simple but straightforward model that employs a simple Archimedean spiral curve to represent the streams, i.e.,
\begin{equation}
r=a+b\theta,
\end{equation}
where $a$ and $b$ are two parameters to determine the curve.

We assumed that the optical ($\lambda=0.3\mu$m band) surface brightness of radiation from the streams is uniformly distributed. We fixed the surface brightness of the triple disks and adjusted the surface brightness values of the stream so that the luminosity we obtain after integrating the surface brightness of the stream is a proportion (e.g., $0.2$) of the luminosity obtained after integrating the surface brightness of the disks.

We find that the presence of the streams influences the amplitude of the light curve, resulting in a larger emission region and a broader light curve. Despite these changes, the light curve continues to exhibit significant periodic variations, with both the phase and period of the variation remaining unchanged (see Fig.~\ref{fig:compare}).

We note here that the existence of streams also affects the overall SED of the system. Specifically, emission from shock-heated gas within these streams can contribute to the UV and optical bands, potentially partially filling the characteristic spectral ``notch'' created by the central cavity \citep[e.g.,][]{Farris15, Cocchiararo24}. However, as indicated by recent 3D simulations, the exact extent to which the streams fill this gap is highly sensitive to binary parameters, orbital phase, and gas thermodynamics, meaning the deficit may not be completely washed out \citep{Franchini24, Cocchiararo24}. Accurately quantifying these exact spectral modifications requires comprehensive radiation-hydrodynamic modeling. Considering our focus is primarily on the geometric and kinematic imprints of the binary on the microlensing light curves, a more detailed discussion of the SED shape variations due to the above uncertainties is beyond the scope of this paper.

\section{Conclusions }
\label{sec:con}
In this study we have explored the microlensing signatures of active SMBBH systems with various mass ratios, accretion rates, and other physical parameters, with a focus on their SEDs and light curves across optical, UV, and X-ray bands. Our key findings are outlined below:

\begin{itemize}

\item {Periodic microlensing signatures:} The orbital dynamics of SMBBHs and their associated mini-disks induce periodic fluctuations in microlensing light curves. The dominant period of these fluctuations is influenced by the binary's mass ratio ($q$): For near-equal mass binaries ($q\approx1$), variations occur at half the orbital period ($P_{\rm orb}/2$) due to both mini-disks alternately intersecting caustic structures. For low mass ratios ($q \lesssim 0.5$), variations align with $P_{\rm orb}$, primarily driven by the secondary mini-disk, especially if it is efficiently accreting, while the primary mini-disk's impact is minimal when the trajectory avoids the caustic-crossing region.

\item {Wavelength-dependent signatures:} Light curves across different bands exhibit the same period and phase. However, the amplitude of microlensing variations is more pronounced in the X-ray band compared to optical bands, due to the more compact nature of the X-ray emitting corona (less than 20 gravitational radii, $r_g$). In UV/optical bands, the half-light radius of mini-disks increases with wavelength, resulting in broader caustic-crossing events and diminished magnification at longer wavelengths. Although the emission regions for X-ray and UV radiation (approximately at a wavelength of $\lambda = 0.1\,\mu$m) are similar in size, the emission distributions from the two mini-disks differ, leading to distinct peak amplitudes. Multiband, multi-epoch monitoring can resolve these differences, aiding in constraining disk sizes and emission distributions.

\item {Parameter constraints:} 
The simple triple-disk architecture of SMBBH systems results in a unique and characteristic flux deficit in the UV/optical SEDs, offering a valuable diagnostic baseline. The signature is influenced by the binary separation, mass ratio, and accretion rates of the binary components. Combining microlensing light curves with SED modeling helps resolve parameter degeneracies (e.g., mass ratio, separation, and accretion rates). While microlensing light curves can impose strong constraints on the periodic motion and radiation sizes of SMBBH-triple disk systems, they alone cannot fully determine system parameters due to degeneracies. SEDs provide insights into SMBH masses and the accretion rates of various disk components. It is crucial to integrate microlensing light curve data with SED observations in order to break parameter degeneracies and accurately characterize SMBBH systems.
However, we note that in highly dynamic environments, cavity streams, shock heating, or radiatively inefficient accretion states may modify and/or partially fill this spectral gap or alter the high-energy emission \citep[e.g.,][]{Farris15, Cocchiararo24, Tiede2025}. A much more complicated model that exclusively considers these complexities is required to infer binary properties from the observed SEDs, which can be challenging.
\end{itemize}

These findings underscore that high-cadence microlensing monitoring of lensed quasars, especially across multiple wavebands, offers a robust method for identifying SMBBH candidates and constraining their properties.
To place these findings in an observational context, we estimated the expected detection rates following the demographic frameworks of \citet{Haiman2009} and \citet{Xin2021}. The fraction of active quasars hosting a binary at a given orbital period ($P_{\rm orb}$), $f_{\rm bin}$, is broadly determined by the ratio of the binary residence time at that separation to the typical quasar lifetime ($\tau_Q \sim 10^7$\,yr). For subparsec binaries with periods of months to a few years, the residence times are typically $\tau_{\rm res} \sim 10^4 - 10^5$\,yr, yielding a binary fraction ($f_{\rm bin}$) on the order of $\sim 10^{-3} - 10^{-2}$. Combined with typical strong-lensing optical depths, this implies that out of the $\sim 10^4$ lensed active galactic nuclei expected to be discovered by the \textit{Vera C. Rubin} Observatory Legacy Survey of Space and Time (LSST), approximately $10$ to $100$ will host close SMBBHs with resolvable macro-images. Detecting periodic microlensing oscillations from this subset is intrinsically unlikely, as it strongly favors short periods ($P_{\rm orb} \lesssim 5$\,yr) and specific caustic-crossing geometries, potentially yielding $\sim \mathcal{O}(1)$ robust candidates per wide-field survey. In contrast, the disk size-wavelength diagnostic applies to wider binaries with longer residence times and could potentially be used to identify $\sim \mathcal{O}(10)$ systems with next-generation facilities. Future high-cadence time-domain surveys like LSST make identifying these unique microlensing signatures a realistic and exciting observational prospect despite their rarity.

\begin{acknowledgements}
This work is partly supported by the National Key Research and Development Program of China (grant nos. 2020YFC2201400, 2022YFC2205201), the National SKA Program of China (grant no. 2020SKA0120101), the National Natural Science Foundation of China (grant nos. 12273050), the Strategic Priority Research Program of the Chinese Academy of Sciences (grant no. XDB0550302), the National Astronomical Observatory of China (grant no. E4TG660101), and the China Manned Space Program with grant no. CMS-CSST-2025-A07. This work was performed in part at the Aspen Center for Physics, which is supported by National Science Foundation grant PHY-2210452.
\end{acknowledgements}

\bibliographystyle{aa} 
\bibliography{ref.bib} 

\end{document}